\documentclass[aps,prb,twocolumn,groupedaddress,showpacs]{revtex4}
\usepackage{graphicx}
\usepackage{amsmath}
\usepackage{amsfonts}
\usepackage{amssymb}

\providecommand{\U}[1]{\protect\rule{.1in}{.1in}}

\begin{document}

\title{Noise and dissipation in magnetoelectronic nanostructures}
\author{J\o rn Foros}
\affiliation{Department of Physics, Norwegian University of Science and Technology, 7491
Trondheim, Norway}
\author{Arne Brataas}
\affiliation{Department of Physics, Norwegian University of Science and Technology, 7491
Trondheim, Norway}
\author{Gerrit E. W. Bauer}
\affiliation{Kavli Institute of NanoScience, Delft University of Technology, 2628 CJ
Delft, The Netherlands}
\author{Yaroslav Tserkovnyak}
\affiliation{Department of Physics and Astronomy, University of California, Los Angeles,
California 90095, USA}
\date{\today}

\begin{abstract}
We study the coupled current and magnetization noise in magnetic
nanostructures by magnetoelectronic circuit theory. Spin current
fluctuations, which depend on the magnetic configuration, are found to be an
important source of magnetization noise and damping in thinly layered
systems. The enhanced magnetization fluctuations in spin valves can be
directly measured by their effect on the resistance noise.
\end{abstract}

\maketitle

\section{Introduction}

New functionalities can be realized by integrating ferromagnetic elements
into electronic circuits and devices. The interplay between magnetism and
electric currents in these structures is utilized by the giant
magnetoresistance (GMR), the operating principle of the read heads in modern
magnetic hard disk drives. Considerable progress has been made in improving
magnetic random access memories.\cite{SPRAM} Efforts to further miniaturize
and improve the performance of magnetoelectronic devices are ongoing in
academic and corporate laboratories. Low power consumption and noise levels
are essential. In spite of the technological relevance, a comprehensive
understanding of coupled current and magnetization noise and the related
energy dissipation in nanoscale magnetoelectronic circuits is lacking.

From the early studies of Johnson\cite{Johnson} and Nyquist,\cite{Nyquist}
we know that the equilibrium voltage noise power in conductors is
proportional to the electric resistance. This relation between the
equilibrium noise and the out-of-equilibrium energy dissipation is a
standard example of the fluctuation-dissipation theorem (FDT).\cite%
{fluct-diss,Landau1} In recent years, important advances have been made in
the understanding of electronic equilibrium (thermal) and non-equilibrium
(shot) noise in mesoscopic conductors.\cite{Blanter-review}

The electron spin plays an important role in electrical noise phenomena in
magnetic multilayers. In early theoretical studies\cite%
{Bulka1998,Tserkovnyak-prb2001,Mishchenko,Lamacraft,Belzig,Zareyan} of
charge and spin-polarized current noise in such systems, magnetizations were
assumed to be static. However, the magnetization itself fluctuates as well.
Thermal fluctuations of the magnetization vector in isolated single-domain
ferromagnets have been analyzed by Brown,\cite{Brown} who introduced a
stochastic Langevin field acting on the magnetization to account for thermal
agitation. His proof that this field's (white-noise) correlator is
proportional to the magnetization damping\ (see below) is another
manifestation of the FDT.\cite{Smith,Safonov} The stochastic field can be
introduced into the spatiotemporal equation of motion for the magnetization
(Landau-Lifshitz-Gilbert equation), affecting, e.g., current-driven
magnetization dynamics and reversal.\cite%
{Wernsdorfer,Koch,Myers-prl2002,Li-prb2004}

A moving magnetization vector in ferromagnets undergoes viscous damping that
relaxes the magnetization toward the lowest (free-)energy configuration.
This process is in practice well described by a phenomenological damping
constant, introduced by Gilbert.\cite{Gilbert,Gilbert2} Despite some
progress,\cite%
{KorenmanPrange1972,Bhagat1974,Suhl1998,Kohno,Skadsem,gilmore2007} a
rigorous quantitative understanding of the magnetic damping in
transition-metal ferromagnets has not yet been achieved. The theory of the
enhanced Gilbert damping in ferromagnets in good electrical contact with a
conducting environment is in a better shape. The loss of angular momentum
due to spin current pumping into the environment agrees with the Gilbert
phenomenology,\cite{prl88,Tserkovnyakreview} and experiment and theory
addressing the additional damping agree well with each other.\cite%
{Tserkovnyakreview}

The electronic and magnetic fluctuations in magnetoelectronic structures are
intimately coupled to each other.\cite{forosPRL,Rebei} For example, the
magnetization noise in ferromagnetic films in good electric contact with
normal metals has been predicted to increase due to spin current
fluctuations: The spin current components polarized perpendicularly to the
magnetization are absorbed at the interface, leading to a fluctuating
spin-transfer torque\cite{Slonczewski1,Berger,Myers,Katine} that induces
additional magnetization noise. This noise is related to the excess Gilbert
damping caused by the angular momentum loss due to spin pumping, in
accordance with the FDT.

Here we investigate the interplay between\ (zero frequency) current and
magnetization noise in multilayers of alternating magnetic and non-magnetic
films. We take advantage of the FDT to relate the equilibrium electric
(current and voltage) and magnetic (magnetization and field) noise to the
corresponding dissipation of energy. We start by reviewing the noise in a
single monodomain ferromagnet sandwiched by normal metals, including
technical details that were omitted in Ref. \onlinecite{forosPRL}. Both
thermal equilibrium (Johnson-Nyquist) current noise and nonequilibrium shot
noise are taken into account. Next, we consider spin valves, \textit{i.e}.,
two ferromagnetic films separated by a normal metal spacer.\cite{forosPRB}
We consider both a symmetric structure in which both layers fluctuate, as
well as an asymmetric one, in which one layer is assumed fixed.
Magnetoelectronic circuit theory\cite{prl84,epjb22,Brataas-review} is used
to calculate the charge and spin current fluctuations. The resulting
enhanced magnetization noise and Gilbert damping in principle are tensors
that depend on the magnetic configuration.

Spin valves provide an opportunity to indirectly measure magnetization noise
via resistance fluctuations, which are manifested by voltage noise for a
current-biased system or current noise for a voltage-biased system.\cite%
{Smith1,Covington} This offers an experimental test of our theory. We obtain
analytical expressions for the magnetic contribution to the induced electric
noise for different magnetic configurations. The noise is of potential
importance for the performance of spin valve read heads.\cite{Smith1} For
symmetric structures in which both layers fluctuate, dynamic cross talk
between the layers becomes important, causing a possibly large difference in
noise level between the parallel and antiparallel magnetic configurations.
Our results for these spin valves include previously presented findings as a
limiting case.\cite{forosPRB} After the completion of this
work,\cite{forosPhD} it was shown that spin-valves in equilibrium also exhibit
colored voltage fluctuations caused by spin pumping of the moving
magnetizations.\cite{Xiao} 

The paper is organized as follows. We begin by reviewing the
fluctuation-dissipation theorem, applied to magnetic systems. In section
III, the noise properties of a single ferromagnetic thin film sandwiched by
normal metals is worked out in detail, emphasizing the relation of the noise
to the damping. In section IV, we consider current noise, magnetization
noise, and magnetization damping in spin valves, and use the results to
calculate the resistance noise induced by GMR. In section V we summarize our
conclusions.

\section{Fluctuation-dissipation theorem}

The fluctuation-dissipation theorem (FDT) relates the spontaneous
time-dependent changes of an observable of a given system in thermal
equilibrium its linear response to an external perturbation that couples to
that observable. For example, in an electric conductor the spontaneous
fluctuations in the electric current are proportional to the dissipative
(real) part of the conductivity, \textit{i.e}., the response function to an
applied electric field.\cite{Johnson,Nyquist} Similarly, the equilibrium
fluctuations of the magnetization vector in a ferromagnet are proportional
to the dissipative part of the magnetic susceptibility, \textit{i.e.},
imaginary part of the response function to an applied magnetic field. In the
following, we briefly recapitulate this FDT for magnetic systems.

Sufficiently below the Curie temperature, changes in the modulus of the
magnetization are energetically costly and may be disregarded. For
sufficiently small magnetic structures spin waves freeze out of the problem.
Hence, a small ferromagnetic particle or thin film is well described in
terms of a single magnetization vector $M_{s}\mathbf{m}$, where $M_{s}$ is
the magnitude of the magnetization and $\mathbf{m}$ a unit vector
(\textquotedblleft macrospin\textquotedblright\ model). The time-dependent
equilibrium fluctuations of the magnetization are characterized by the
autocorrelation function $\langle \delta {m}_{i}(t)\delta {m}_{j}(t^{\prime
})\rangle $, where $\delta {m}_{i}(t)={m}_{i}(t)-\langle {m}_{i}(t)\rangle $
are transverse fluctuations. Here the brackets denote statistical averaging
at equilibrium, and $i$ and $j$ denote Cartesian components perpendicular to
the equilibrium/average magnetization direction. The classical FDT states
that these fluctuations are related to the magnetic susceptibility: 
\begin{align}
\langle \delta m_{i}(t)\delta m_{j}(t^{\prime })\rangle =& \frac{k_{B}T}{%
2\pi M_{s}V}\int d\omega e^{-i\omega (t-t^{\prime })}  \notag \\
& \times \frac{\chi _{ij}(\omega )-\chi _{ji}^{\ast }(\omega )}{i\omega },
\label{FDT}
\end{align}%
where $T$ is the temperature, $V$ the volume of the ferromagnet, and $\chi
_{ij}(\omega )$ the $ij$-component of the transverse magnetic susceptibility
at frequency $\omega $. The latter is the linear (causal) response function
that describes the changes of the magnetization, $\Delta m_{i}(t)$, caused
by an external driving field $\mathbf{H}^{(\mathrm{dr})}(t)$: 
\begin{equation}
\Delta m_{i}(t)=\sum_{j}\int dt^{\prime }\chi _{ij}(t-t^{\prime }){H}_{j}^{(%
\mathrm{dr})}(t^{\prime }).  \label{magsusc}
\end{equation}%
An alternative form of the FDT that turns out useful in the course of this
paper can be derived by introducing a stochastic magnetic field $\mathbf{h}%
^{(0)}(t)$ with zero mean. This field effectively represents the coupling of
the magnetization to the dissipative degrees of freedom, and is viewed as
the cause of the thermal fluctuations $\delta \mathbf{m}(t)$. The
microscopic origin of $\mathbf{h}^{(0)}(t)$ does not concern us here, but it
might, e.g., represent thermally excited phonons that deform the crystal
anisotropy fields. From Eq.~(\ref{magsusc}) it follows that $\delta
m_{i}(\omega )=\sum_{j}\chi _{ij}(\omega ){h}_{j}^{(0)}(\omega )$ in
frequency domain. Inverting this relation, the correlator of the stochastic
field has to obey the relation 
\begin{align}
\langle h_{i}^{(0)}(t)h_{j}^{(0)}(t^{\prime })\rangle =& \frac{k_{B}T}{2\pi
M_{s}V}\int d\omega e^{-i\omega (t-t^{\prime })}  \notag \\
& \times \frac{\lbrack \chi _{ji}^{-1}(\omega )]^{\ast }-\chi
_{ij}^{-1}(\omega )}{i\omega },  \label{FDT2}
\end{align}%
where $\chi _{ij}^{-1}(\omega )$ is the $ij$-component of the Fourier
transformed inverse susceptibility.

\section{Single ferromagnet}

\label{SingleFerromagnet}

The magnetization dynamics of an isolated single-domain ferromagnet is well
described by the Landau-Lifshitz-Gilbert (LLG) equation\cite{Landau,Gilbert} 
\begin{equation}
\frac{d\mathbf{m}}{dt}=-\gamma _{0}\mathbf{m}\times \mathbf{H}_{\mathrm{eff}%
}+\alpha _{0}\mathbf{m}\times \frac{d\mathbf{m}}{dt},  \label{LLG}
\end{equation}%
where $\gamma _{0}$ is the gyromagnetic ratio, $\mathbf{H}_{\mathrm{eff}}$
the effective magnetic field, and $\alpha _{0}$ the Gilbert damping
constant. The effective field has contributions due to crystal and form
anisotropies, as well as externally applied magnetic fields. By linearizing
this LLG equation we can evaluate the magnetic susceptibility and the
equilibrium magnetization noise. The average equilibrium direction of the
magnetization is aligned with $\mathbf{H}_{\mathrm{eff}}$ to minimize the
energy: $\mathbf{m}_{0}=\mathbf{H}_{\mathrm{eff}}/|\mathbf{H}_{\mathrm{eff}}|
$. A weak external driving field is included by substituting $\mathbf{H}_{%
\mathrm{eff}}\rightarrow \mathbf{H}_{\mathrm{eff}}+\mathbf{H}^{(\mathrm{dr}%
)}(t).$ In the present model only the component of $\mathbf{H}^{(\mathrm{dr}%
)}$ transverse to the magnetization will solicit a response $\mathbf{m}%
(t)\approx \mathbf{m}_{0}+\Delta \mathbf{m}(t)$ of the magnetization. Here $%
\Delta \mathbf{m}(t)$ is normal to $\mathbf{m}_{0}$. To  lowest order in $%
\Delta \mathbf{m}(t)$, the LLG equation gives the inverse susceptibility
tensor matrix 
\begin{equation}
{\chi }^{-1}=\frac{1}{\gamma _{0}}\left[ 
\begin{array}{cc}
\gamma _{0}\left\vert \mathbf{H}_{\mathrm{eff}}\right\vert -i\omega \alpha
_{0} & i\omega  \\ 
-i\omega  & \gamma _{0}\left\vert \mathbf{H}_{\mathrm{eff}}\right\vert
-i\omega \alpha _{0}%
\end{array}%
\right]   \label{suscmatrix}
\end{equation}%
in the plane normal to $\mathbf{m}_{0}$. A dependence of the effective field 
$\mathbf{H}_{\mathrm{eff}}$ on $\mathbf{m}$  does not affect the noise
properties.

The magnetization noise follows from substituting Eq.~(\ref{suscmatrix})
into Eq.~(\ref{FDT}). The correlator of the stochastic field is obtained
from Eqs.~(\ref{FDT2}) and (\ref{suscmatrix}) and does not depend on the
effective field:\cite{Brown} 
\begin{equation}
\langle h_{i}^{(0)}(t)h_{j}^{(0)}(t^{\prime})\rangle=2k_{B}T\frac{\alpha_{0}%
}{\gamma_{0}M_{s}V}\delta_{ij}\delta(t-t^{\prime}).   \label{h0corr}
\end{equation}
The relation between the equilibrium magnetization fluctuations and the
dissipation in the form of the Gilbert damping is evident.

Up to now we considered a ferromagnet isolated from the outside world. Its
dynamics is altered by embedding into a conducting environment.\cite{prl88}
A ferromagnet with time-dependent magnetization \textquotedblleft
pumps\textquotedblright\ an angular-momentum (spin) current 
\begin{equation}
\mathbf{I}_{s}^{\mathrm{pump}}=\frac{\hbar }{4\pi }\left( \mathrm{Re}%
g^{\uparrow \downarrow }\mathbf{m}\times \frac{d\mathbf{m}}{dt}+\mathrm{Im}%
g^{\uparrow \downarrow }\frac{d\mathbf{m}}{dt}\right) ,  \label{spinpumping1}
\end{equation}%
into an adjacent conductor. Here $g^{\uparrow \downarrow }$ is the
dimensionless transverse spin (\textquotedblleft spin
mixing\textquotedblright\ ) conductance that depends on the interface
transparency between ferromagnet and proximate metal.\cite%
{prl84,epjb22,Brataas-review} When the spin current is efficiently
dissipated in the conductor, thus does not build up a spin accumulation
close to the interface, the loss of angular momentum corresponds to an extra
torque $\gamma \mathbf{I}_{s}^{\mathrm{pump}}/(M_{s}V)$ on the right hand
side of the Eq. (\ref{LLG}). This is equivalent to an increased Gilbert
damping and a modified gyromagnetic ratio:\cite{prl88} 
\begin{equation}
\frac{1}{\gamma _{0}}\rightarrow \frac{1}{\gamma }=\frac{1}{\gamma _{0}}%
\left( 1-\frac{\gamma _{0}\hbar \mathrm{{Im}}g^{\uparrow \downarrow }}{%
4\pi M_{s}V}\right) ,  \label{spinpumping2}
\end{equation}%
\begin{equation}
\alpha _{0}\rightarrow \alpha =\frac{\gamma }{\gamma _{0}}\left( \alpha _{0}+%
\frac{\gamma _{0}\hbar \mathrm{{Re}}g^{\uparrow \downarrow }}{4\pi
M_{s}V}\right) .  \label{spinpumping3}
\end{equation}%
In the strong coupling limit (intermetallic interfaces), $\mathrm{{Im}}%
g^{\uparrow \downarrow }\ll \mathrm{{Re}}g^{\uparrow \downarrow }$ and
we are allowed to disregard the difference between $\gamma $ and $\gamma _{0}
$.

Another term which modifies the magnetization dynamics is the so-called
spin-transfer torque.\cite{Slonczewski1,Berger,Myers,Katine} It is also
proportional to the spin-mixing conductance introduced above\cite%
{prl84,epjb22,Brataas-review} and represented by adding $-\gamma _{0}\mathbf{%
I}_{s,\mathrm{abs}}/(M_{s}V)$ to the right hand side of Eq. (\ref{LLG}).
Here\ $\mathbf{I}_{s,\mathrm{abs}}$ is the spin-polarized current
transversely polarized to the magnetization, which is absorbed by the
ferromagnet on an atomic length scale, thereby transferring its angular
momentum to the magnetization. Spin pumping and spin-transfer torque are
related by an Onsager reciprocity relation.\cite{TserkovnyakMecklenburg2008}

Recently we have shown\cite{forosPRL} that the magnetization noise in
magnetoelectronic nanostructures can be considerably increased as compared
to an isolated ferromagnet. At elevated temperatures, thermal fluctuations
in the spin current exert a fluctuating torque on the magnetization,
increasing the noise. For a ferromagnet sandwiched by normal metals, the
enhancement of the noise is described by a stochastic field $\mathbf{h}^{(%
\mathrm{th})}(t)$ similar to the intrinsic field $\mathbf{h}^{(0)}(t)$. Its
correlation function reads\cite{forosPRL} 
\begin{equation}
\langle h_{i}^{(\mathrm{th})}(t)h_{j}^{(\mathrm{th})}(t^{\prime})%
\rangle=2k_{B}T\frac{\alpha^{\prime}}{\gamma M_{s}V}\delta_{ij}\delta(t-t^{%
\prime}),   \label{hthcorr}
\end{equation}
where 
\begin{equation}
\alpha^{\prime}=\frac{\gamma\hbar\mathrm{{Re}}g^{\uparrow \downarrow}}{%
4\pi M_{s}V}
\end{equation}
is the enhancement of the Gilbert damping due to spin pumping (see Eq. (\ref%
{spinpumping3})). Assuming that $\mathbf{h}^{(0)}(t)$ and $\mathbf{h}^{(%
\mathrm{th})}(t)$ are statistically independent, the total magnetization
noise is thus given by $\mathbf{h}(t)=\mathbf{h}^{(0)}(t)+\mathbf{h}^{(%
\mathrm{th})}(t)$. We know that the total damping is determined by $%
\alpha=\alpha_{0}+\alpha^{\prime}$, and from Eqs. (\ref{h0corr}) and (\ref%
{hthcorr}) we see that the total noise is related to the total damping, in
agreement with the FDT. Hence, the thermal spin current noise is the
stochastic process related to the enhanced dissipation of energy by spin
pumping. By calculating the noise power we also know the damping and vice
versa. In thin ferromagnetic films, $\alpha^{\prime}$ can be of the same
order or even larger than $\alpha_{0}$.\cite{Tserkovnyakreview} In the
following subsections we will give a detailed derivation of Eq. (\ref%
{hthcorr}). We also evaluate the shot noise contribution to the
magnetization noise, which is important at low temperatures.\cite{forosPRL}
We note here that Eq. (\ref{hthcorr}) may be found also by direct
application of Eq. (\ref{FDT2}) to the LLG equation with spin-pumping
included.

\subsection{Scattering theory}

\label{LB}

\begin{figure}[ptb]
\includegraphics{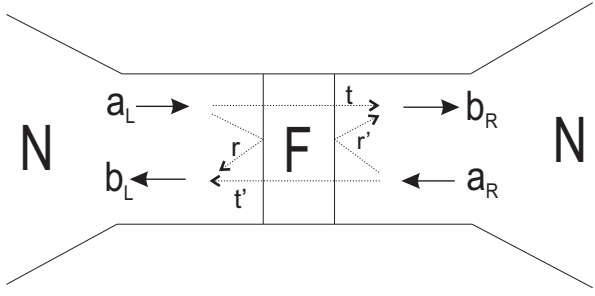} 
\caption{A thin ferromagnetic (F) film is sandwiched by normal metals (N).
The current fluctuations in the system are evaluated in terms of
transmission probabilities for the electron states, with the aid of second
quantized annihilation and creation operators. The operators shown in the
figure are annihilation operators, with the $a$-operators annihilating
electrons moving towards the ferromagnet, and the $b$-operators annihilating
electrons moving away from the ferromagnet. Also shown are the reflection
and transmission matrices $r,r^{\prime},t,t^{\prime}$ (see Eq. (\protect\ref%
{scatteringmatrix})), for simplicity without spin indices.}
\label{system}
\end{figure}

We study a thin ferromagnetic film connected to two normal reservoirs, as
shown in Fig. \ref{system}. The reservoirs are perfect spin sinks and the
ferromagnet is taken to be thicker than the magnetic coherence length $%
\lambda_{c}=\pi/(k_{\uparrow}-k_{\downarrow})$, where $k_{\uparrow
(\downarrow)}$ are spin-dependent Fermi momenta. For transition metals, $%
\lambda_c$ is of the order of monolayers. The normal metals are
characterized by Fermi-Dirac distribution functions $f_{L}$ and $f_{R}$ with
chemical potentials $\mu_{L}$ and $\mu_{R}$, where $L$ and $R$ refer to the
left and right sides at a common temperature $T$. We use the Landauer-B{\"{u}%
}ttiker (LB) scattering theory\cite{Blanter-review} to evaluate the spin
current fluctuations, and the LLG equation to calculate the resulting
magnetization noise.

In the LB approach electron transport is expressed in terms of transmission
probabilities between the electron states on different sides of a scattering
region. Here we interpret the ferromagnetic film as a scatterer that limits
the propagation of electrons between the normal reservoirs. The scattering
properties of the ferromagnet and the bias between the reservoirs determine
the transport properties of the system. The transport channels in the leads
are modelled as ideal electron wave guides in which the transverse and
longitudinal motions are separable. The transport channels at a given energy 
$E$ are then labeled by the discrete mode index for the quantized transverse
motion, by which the continuous wave vector for the longitudinal motion is
fixed. The LB formalism\cite{Buttiker-prb1992,Blanter-review} generalized to
describe spin transport leads to the current operator 
\begin{align}
\hat{I}_{A}^{\alpha\beta}(t) =& \frac{e}{h}\int dEdE^{\prime}e^{
i(E-E^{\prime })t/\hbar}  \notag \\
&\times[a_{A\beta}^{\dagger}(E)a_{A\alpha}(E^{\prime})-b_{A\beta}^{%
\dagger}(E)b_{A\alpha}(E^{\prime})].   \label{currentoperator}
\end{align}
at time $t$ on side $A$ [$=L$(left) or $R$(right)] of the ferromagnetic
film. Here, $\alpha$ and $\beta$ denote components in $2\times2$ spin space. 
$a_{A\alpha}(E)$ and $b_{A\alpha}(E)$ are operators for all transport
channels at energy $E$ that annihilate electrons with spin $\alpha$ in lead
A that move towards and away from the ferromagnet, respectively (see Fig. %
\ref{system}). The $a$-operators are related to the $b$-operators by the
scattering properties of the ferromagnet: 
\begin{equation}
b_{A\alpha}(E)=\sum_{B\beta}s_{AB\alpha\beta}(E)a_{B\beta}(E), 
\label{barelation}
\end{equation}
where $s_{AB\alpha\beta}$ is the scattering matrix for incoming electrons
with spin $\beta$ in lead $B$ $(=L$ or $R)$ scattered to outgoing states in
lead $A$ with spin $\alpha$. The summation is over $B=L,R$ and over spin $%
\beta=\uparrow,\downarrow$. A similar relation holds for the creation
operators. Current conservation implies that the scattering matrix is
unitary. Suppressing spin indices for simplicity (see Fig. \ref{system}) 
\begin{equation}
\left( 
\begin{array}{c}
b_{L} \\ 
b_{R}%
\end{array}
\right) =\left( 
\begin{array}{cc}
r & t^{\prime} \\ 
t & r^{\prime}%
\end{array}
\right) \left( 
\begin{array}{c}
a_{L} \\ 
a_{R}%
\end{array}
\right) ,   \label{scatteringmatrix}
\end{equation}
where $r=s_{LL}$, $r^{\prime}=s_{RR}$, $t=s_{RL}$ and $t^{\prime}=s_{LR}$.
In the following we disregard spin-flip processes in the ferromagnet.
Choosing the spin quantization $z$-axis in the direction of the average
magnetization, this implies that $s_{AB\alpha\beta}=s_{AB\alpha}\delta_{%
\alpha\beta }$.

The outgoing charge and spin currents are given respectively by $%
I_{c,A}(t)=\sum_{\alpha}\hat{I}_{A}^{\alpha\alpha}(t)$ and $\mathbf{I}%
_{s,A}(t)=-(\hbar/2e)\sum_{\alpha\beta}\hat{\mbox{\boldmath $\sigma$}}%
^{\alpha\beta }\hat{I}_{A}^{\beta\alpha}(t)$, where $\hat{\mbox{\boldmath
$\sigma$}}=(\hat{\sigma}_{x},\hat{\sigma}_{y},\hat{\sigma}_{z})$ is the
vector of Pauli matrices. The expectation values for charge and spin
currents are evaluated using the quantum statistical average $\langle
a_{Am\alpha}^{\dagger
}(E)a_{Bn\beta}(E^{\prime})\rangle=\delta_{AB}\delta_{mn}\delta_{\alpha\beta
}\delta(E-E^{\prime})f_{A}(E)$ of the product of one creation and one
annihilation operator, where $m$ and $n$ label the transport channels. The
creation and annihilation operators obey the anticommutation relation 
\begin{equation}
\{a_{Am\alpha}^{\dagger}(E),a_{Bn\beta}(E^{\prime})\}=\delta_{AB}\delta
_{mn}\delta_{\alpha\beta}\delta(E-E^{\prime}),
\end{equation}
whereas the anticommutators of two creation or two annihilation operators
vanish. Similar relations hold for the $b$ operators. The average 
\begin{align}
& \langle a_{Ak\alpha}^{\dagger}(E_{1})a_{Bl\beta}(E_{2})a_{Cm\gamma
}^{\dagger}(E_{3})a_{Dn\delta}(E_{4})\rangle  \notag \\
&-\langle a_{Ak\alpha}^{\dagger}(E_{1})a_{Bl\beta}(E_{2})\rangle\langle
a_{Cm\gamma}^{\dagger}(E_{3})a_{Dn\delta}(E_{4})\rangle  \notag \\
&
=\delta_{AD}\delta_{BC}\delta_{kn}\delta_{lm}\delta_{\alpha\delta}\delta_{%
\beta\gamma}\delta(E_{1}-E_{4})\delta(E_{2}-E_{3})  \notag \\
&\hspace{0.4cm}\times f_{A}(E_{1})[1-f_{B}(E_{2})],   \label{expectation}
\end{align}
where the subscripts $A,B,C,D$ denote leads, $k,l,m,n$ transport channels,
and $\alpha,\beta,\gamma,\delta$ spin, is needed for the calculation of the
current fluctuations. We also need the identity 
\begin{equation}
\sum_{CD}\mathrm{Tr}(s_{AC\alpha}^{\dagger}s_{AD\beta}s_{BD\beta}^{\dagger
}s_{BC\alpha})=\delta_{AB}M_{A} \,,  \label{trace}
\end{equation}
which follows from the unitarity of the scattering matrix. Here the trace is
over the space of the transport channels, and $M_{A}$ is the number of
transverse channels in lead $A$, all at a given energy.

The charge and spin current correlation functions read 
\begin{equation}
S_{c,AB}(t-t^{\prime})=\langle\delta I_{c,A}(t)\delta I_{c,B}(t^{\prime
})\rangle
\end{equation}
and 
\begin{equation}
S_{ij,AB}(t-t^{\prime})=\langle\delta I_{s_{i},A}(t)\delta
I_{s_{j},B}(t^{\prime})\rangle,
\end{equation}
where $\delta I_{c,A}(t)=I_{c,A}(t)-\langle I_{c,A}(t)\rangle$ denotes the
deviation of the charge current from its average value in lead $A$ at time $t
$, and $\delta I_{s_{i},A}(t)$ is the deviation of the vector component $i$ (%
$i=x,y$ or $z$) of the spin current. We are interested mainly in the
low-frequency noise, \textit{i.e.}, the time integrated value of the
correlation functions: 
\begin{equation}
S_{c,AB}(\omega=0)=\int d(t-t^{\prime})S_{c,AB}(t-t^{\prime}).
\end{equation}
Two fundamentally different types of current noise have to be distinguished:
Thermal (equilibrium) noise and (non-equilibrium) shot noise. In general,
the total noise is not simply a linear combination of both types.
Nevertheless, it is convenient to treat the two noise sources independently,
by separately investigating the noise of an unbiased system at finite
temperatures in Sec. \ref{TCN} and the shot noise under an applied bias at
zero temperature in Sec. \ref{SN}.

\subsection{Thermal current noise\label{TCN}}

At equilibrium $f_{L}=f_{R}=f$, and the average current vanishes. However,
at finite temperatures, the occupation numbers of the electron channels
incident on the sample fluctuate in time and so does the current. Using
Eqs.~(\ref{currentoperator}), (\ref{barelation}), (\ref{expectation}), (\ref%
{trace}) and $f(1-f)=k_{B}T(-\partial f/\partial E)$, we recover the
well-known Johnson-Nyquist noise 
\begin{equation}
S_{c,AA}^{(\mathrm{th})}(\omega=0)=\frac{2e^{2}}{h}k_{B}T(g^{\uparrow
}+g^{\downarrow})   \label{JohnsonNyquist}
\end{equation}
in the zero-frequency limit. Here $g^{\alpha}=\mathrm{Tr}(1-r_{\alpha}^{%
\dagger}r_{\alpha}),$ where the trace indicates again a summation over
transport channels, is the spin-dependent dimensionless conductance of the
ferromagnet, to be evaluated at the Fermi energy. The superscript $(\mathrm{%
th})$ emphasizes that the fluctuations are caused by thermal agitation. The
result for $S_{c,AB}^{(\mathrm{th})}(\omega=0)$, where $B\neq A$, differs
from the above expression only by a minus sign, since current direction is
defined positive towards the ferromagnet on both sides, and charge current
is conserved. The Johnson-Nyquist noise, Eq. (\ref{JohnsonNyquist}), is a
manifestation of the FDT, since it relates the equilibrium current noise to
the dissipation of energy prarameterized by the conductance.

The thermal \emph{spin current} noise can be obtained in a similar way. At
zero frequency 
\begin{align}
S_{ij,AB}^{(\mathrm{th})}(0) =& \frac{\hbar k_{B}T}{8\pi}\sum_{\alpha\beta
}\sigma_{i}^{\alpha\beta}\sigma_{j}^{\beta\alpha}  \notag \\
&\hspace{-0.5cm}\times\mathrm{Tr}[2\delta_{AB}-s_{BA\alpha}^{\dagger}s_{BA%
\beta}-s_{AB\beta}^{\dagger}s_{AB\alpha }],   \label{spincurrentnoise}
\end{align}
where the scattering matrices should again be evaluated at the Fermi energy.
The noise power of the $z$ component (polarized parallel to the
magnetization) of the spin current 
\begin{equation}
S_{zz,AA}^{(\mathrm{th})}=\frac{\hbar}{4\pi}k_{B}T(g^{\uparrow}+g^{%
\downarrow })
\end{equation}
differs from the charge current noise only by the squared conversion factor, 
$(\hbar/2e)^{2}$, from charge to spin currents. The transverse (polarized
perpendicular to the magnetization) spin-current components fluctuate as 
\begin{equation}
S_{xx,AA}^{(\mathrm{th})}=S_{yy,AA}^{(\mathrm{th})}=\frac{\hbar}{4\pi}%
k_{B}T(g_{A}^{\uparrow\downarrow}+g_{A}^{\downarrow\uparrow}).
\end{equation}
The \textquotedblleft spin mixing\textquotedblright\ conductances $%
g_{L}^{\uparrow\downarrow}=\mathrm{Tr}[1-r_{\uparrow}(r_{\downarrow
})^{\dagger}]=(g_{L}^{\downarrow\uparrow})^{\ast}$ and $g_{R}^{\uparrow
\downarrow}=\mathrm{Tr}[1-r_{\uparrow}^{\prime}(r_{\downarrow}^{\prime
})^{\dagger}]=(g_{R}^{\downarrow\uparrow})^{\ast}$ parametrize the
absorbtivity of the ferromagnetic interfaces for transverse-polarized spin
currents. We see that also the spin-current noise obeys the FDT, since the
spin-current correlators are proportional to the conductances for the
respective spin current components.

The cross correlation $S_{zz,LR}^{(\mathrm{th})}=-S_{zz,LL}^{(\mathrm{th})}$
reflects conservation of the longitudinal spin current in the ferromagnet,
since spin-flip scattering is disregarded. On the other hand, $S_{xx,LR}^{(%
\mathrm{th})}=S_{yy,LR}^{(\mathrm{th})}=0$, because the transverse spin
current is absorbed at the interfaces to a ferromagnet thicker than the
magnetic coherence length.

\subsection{Shot noise\label{SN}}

Shot noise of the electronic charge current is an out-of-equilibrium
phenomenon proportional to the current bias. Shot noise is due to the
discreteness of the electron charge, and the probabilistic incidence of
electrons on the scatterer/resistor. Let $\mu_{L}-\mu_{R}=eU$ with $U$ the
applied voltage, and take the temperature to be zero. We are here only
concerned with the current fluctuations, although in this case also the
average charge current is nonzero. The average spin current accompanying the
average charge current does not exert a torque on a single ferromagnet,
since the spin current is polarized along the direction of magnetization.
From Eqs. (\ref{currentoperator}), (\ref{barelation}), (\ref{expectation}),\
and making use of the zero temperature relations $f_{A}(1-f_{A})=0$ and $%
\int dE(f_{L}-f_{R})^{2}=e|U|$, we reproduce the well-known charge shot
noise expression\cite{Blanter-review} 
\begin{equation}
S_{c,AA}^{(\mathrm{sh})}(0)=\frac{e^{3}}{h}|U|[\mathrm{Tr}(r_{\uparrow
}^{\dagger}r_{\uparrow}t_{\uparrow}^{\dagger}t_{\uparrow})+\mathrm{Tr}%
(r_{\downarrow}^{\dagger}r_{\downarrow}t_{\downarrow}^{\dagger}t_{\downarrow
})]   \label{chargeshotnoise}
\end{equation}
Again, the scattering matrices should be evaluated at the Fermi energy, and
the superscript $(\mathrm{sh})$ emphasizes that this is shot noise. $%
S_{c,AB}^{(\mathrm{sh})}(0)=-S_{c,AA}^{(\mathrm{sh})}(0)$, where $B\neq A$.
The spin current shot noise power is 
\begin{align}
S_{ij,AB}^{(\mathrm{sh})}(0) & =\frac{\hbar}{8\pi}\sum_{\alpha\beta}\hat{%
\sigma}_{i}^{\alpha\beta}\hat{\sigma}_{j}^{\beta\alpha}\int dE\sum
_{CD}f_{C}(1-f_{D})  \notag \\
& \times\mathrm{Tr}[s_{AC\alpha}^{\dagger}s_{AD\beta}s_{BD\beta}^{\dagger
}s_{BC\alpha}].   \label{spinshotnoise}
\end{align}
From this we find, $S_{zz,LR}^{(\mathrm{sh})}=-S_{zz,LL}^{(\mathrm{sh})}$
and $S_{xx,LR}^{(\mathrm{th})}=S_{yy,LR}^{(\mathrm{th})}=0$, which hold for
the same reasons as for the thermal noise.

\subsection{Magnetization noise and damping \label{recipe}}

The absorption of fluctuating transverse spin currents at the ferromagnet's
interfaces implies a fluctuating spin-transfer torque on the magnetization.
The resulting increment of the magnetization noise can be calculated using
Eq. (\ref{LLG}), which by conservation of angular momentum is modified by
the spin torque $-\gamma _{0}\mathbf{I}_{s,\mathrm{abs}}\left( t\right)
/(M_{s}V)$. Here $\mathbf{I}_{s,\mathrm{abs}}=\mathbf{I}_{s,L}+\mathbf{I}%
_{s,R}$ is the (instantaneously) absorbed spin current. (Recall that on both
sides of the ferromagnet, positive current direction is defined towards the
magnet.) Since $\mathbf{I}_{s,\mathrm{abs}}$ is perpendicular to $\mathbf{m}$%
, we may in general write $\mathbf{I}_{s,\mathrm{abs}}=-\mathbf{m}\times
\lbrack \mathbf{m}\times \mathbf{I}_{s,\mathrm{abs}}]$, such that the
modified stochastic LLG equation reads 
\begin{align}
\frac{d\mathbf{m}}{dt}& =-\gamma _{0}\mathbf{m}\times \lbrack \mathbf{H}_{%
\mathrm{eff}}+\mathbf{h}^{(0)}(t)]+\alpha _{0}\mathbf{m}\times \frac{d%
\mathbf{m}}{dt}  \notag \\
& +\frac{\gamma _{0}}{M_{s}V}\mathbf{m}\times \lbrack \mathbf{m}\times 
\mathbf{I}_{s_{\mathrm{abs}}}].  \label{LLG2}
\end{align}%
For the single ferromagnetic scatterer $\langle \mathbf{I}_{s,\mathrm{abs}%
}\rangle =0$, but $\delta \mathbf{I}_{s,\mathrm{abs}}(t)\neq 0$. We can thus
define $\mathbf{h}(t)=-1/(M_{s}V)\mathbf{m}\times \delta \mathbf{I}_{s}(t)$
to be a stochastic "magnetic" field that takes into account the (thermal or
shot) spin current noise that comes in addition to the intrinsic noise field 
$\mathbf{h}^{\mathrm{(0)}}(t)$. The correlators of the field 
\begin{equation}
\langle h_{i}(t)h_{i}(t^{\prime })\rangle =\frac{1}{M_{s}^{2}V^{2}}%
\sum_{AB}S_{jj,AB}(t-t^{\prime })
\end{equation}%
and 
\begin{equation}
\langle h_{i}(t)h_{j}(t^{\prime })\rangle =-\frac{1}{M_{s}^{2}V^{2}}%
\sum_{AB}S_{ji,AB}(t-t^{\prime })
\end{equation}%
for $i,j=x,y$; $i\neq j$ are directly obtained from the current noise. $%
\mathbf{h}(t)$ per definition has no component parallel to the
magnetization. In the limit that the current noise is `white' on the
relevant energy scales (temperature, applied voltage, and exchange
splitting), we can approximate $S_{ij,AB}(t-t^{\prime })\approx
S_{ij,AB}(\omega =0)\delta (t-t^{\prime })$. Using Eq. (\ref%
{spincurrentnoise}) we then find the already advertised result 
\begin{equation}
\langle h_{i}^{\mathrm{(th)}}(t)h_{j}^{\mathrm{(th)}}(t^{\prime })\rangle
=2k_{B}T\frac{\alpha ^{\prime }}{\gamma _{0}M_{s}V}\delta _{ij}\delta
(t-t^{\prime }),  \label{advertisedResult}
\end{equation}%
for the thermally (th) induced stochastic field. Here $\alpha ^{\prime
}=\gamma _{0}\hbar \mathrm{Re}(g_{L}^{\uparrow \downarrow }+g_{R}^{\uparrow
\downarrow })/(4\pi M_{s}V)$ is the spin-pumping enhancement of the Gilbert
damping constant. This result is in agreement with the FDT [Eq.~(\ref{FDT2}%
)] with a total Gilbert damping $\alpha =\alpha _{0}+\alpha ^{\prime }$.

Using Eq. (\ref{spinshotnoise}) and the unitarity of the scattering matrix
we find for the stochastic field generated by the shot noise $\mathbf{h}^{%
\mathrm{(sh)}}:$ 
\begin{align}
\langle h_{i}^{\mathrm{(sh)}}(t)h_{j}^{\mathrm{(sh)}}(t^{\prime})\rangle =&%
\frac{\hbar}{4\pi}\frac{e|U|}{M_{s}^{2}V^{2}}\delta_{ij}\delta(t-t^{\prime
})[\mathrm{Tr}(r_{\uparrow}r_{\uparrow}^{\dagger}t_{\downarrow}^{\prime
}t_{\downarrow}^{\prime\dagger})  \notag \\
&+ \mathrm{Tr}(r_{\downarrow}^{\prime}r_{\downarrow}^{\prime\dagger
}t_{\uparrow}t_{\uparrow}^{\dagger})]\,.
\end{align}
For a simple Stoner model it can be shown that for typical experimental
voltage drops in nanoscale metallic spin valves, $\mathbf{h}^{\mathrm{(sh)}}$
can dominate $\mathbf{h}^{\mathrm{(th)}}$ at temperatures of the order of 10 K.\cite{forosPRL} In the following section we concentrate on room
temperature, at which shot noise may be disregarded.

\section{Spin valves}

\begin{figure}[ptb]
\includegraphics[width=0.45\textwidth]{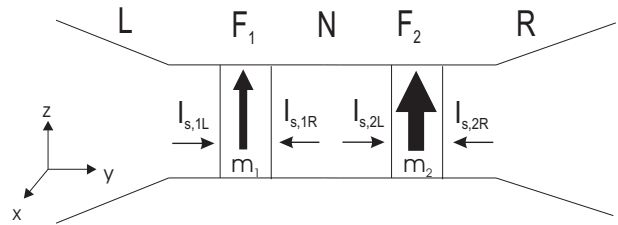} 
\caption{A spin valve with two ferromagnets $F_{1}$ and $F_{2}$ with unit
magnetization vectors $\mathbf{m}_{1}$ and $\mathbf{m}_{2}$, here shown in
the parallel (P) configuration $\mathbf{m}_{1}=\mathbf{m}_{2}=\mathbf{z}$.
The magnetization of $F_{2}$ is fixed. The currents in the system are
evaluated by magnetoelectronic circuit theory on the normal side of the
interfaces, with positive directions defined by the arrows.}
\label{spinvalve}
\end{figure}

We now proceed to consider the noise properties of spin valve nanopillars, 
\textit{i.e}., layered structures consisting of two ferromagnets $F_{1}$ and 
$F_{2}$ with respective unit magnetization vectors $\mathbf{m}_{1}$ and $%
\mathbf{m}_{2}$ that are separated by a thin normal metal spacer $N$, as
sketched in Fig. \ref{spinvalve}. We first assume that $F_{2}$ is highly
coercive, such that the fluctuations of it's magnetization vector are small.
Such a `pinning' is routinely achieved in spin valves, \textit{e.g}., by
`exchange biasing.' We relax this condition in Sec \ref{TIFM}.

The magnetization noise of the free layer $F_{1}$ is caused by intrinsic
processes as well as by fluctuating spin currents in the neighbouring normal
metals. The latter source is affected by the presence of the second
ferromagnet. Magnetoelectronic circuit theory\cite%
{prl84,epjb22,Brataas-review} enables us to compute the current fluctuations
and thus the magnetizations noise of composite structures such as spin
valves.

Fluctuations of $\mathbf{m}_{1}$ cause an easily measurable electrical
noise, since the resistance of a spin valves depends on the relative
orientation of the magnetizations (GMR). Resistance noise is also
interesting from a technological point of view, since it affects the
sensitivity of spin valve read heads in magnetic storage devices.

In the following, we briefly explain the spin current noise calculation by
magnetoelectronic circuit theory. The stochastic field that acts on the free
layer $F_{1}$ and the related Gilbert damping are found for different
magnetic configurations. Using the LLG equation, we then calculate the
fluctuations of the magnetization vector and the resulting resistance noise.
We finish this section by considering spin valves in which both ferromagnets
are identically susceptible to fluctuations.

\subsection{Circuit theory}

Magnetoelectronic circuit theory\cite{prl84,epjb22,Brataas-review} is a tool
to determine transport properties of magnetoelectronic heterostructures such
as the spin valve shown in Fig. \ref{spinvalve}. It is based on the division
of a given structure into resistive elements (scatterers), nodes (low
resistance interconnects), and reservoirs (voltage sources). The current
through local resistors is calculated by LB scattering theory, which
requires that nodes and reservoirs are characterized by (semiclassical)
distribution functions. Here we take the ferromagnetic inserts as
scatterers, the central normal metal layer as a node, and the outer normal
metals $L$ (left) and $R$ (right) as large reservoirs. The reservoirs are in
thermal equilibrium, and hence characterized by Fermi-Dirac distribution
functions $f_{L}=f(E-\mu_{L})$ and $f_{R}=f(E-\mu_{R})$, where $\mu_{L}$ and 
$\mu_{R}$ are the respective chemical potentials. Depending on the relative
orientation of the magnetization vectors $\mathbf{m}_{1}$ and $\mathbf{m}_{2}
$, there can be a non-equilibrium accumulation of spins on the normal metal
node, thus characterized by a scalar (charge) distribution function $f_{cN}$%
, and a vector spin distribution function $\mathbf{f}_{sN}$. $f_{cN}$ and $%
\mathbf{f}_{sN}$ form the distribution matrix $\hat{f}_{N}=\hat{1}f_{cN}+%
\hat{\boldsymbol{\sigma}}\cdot\mathbf{f}_{sN}$ in $2\times2$ spin space. As
before, the ferromagnets are thicker than $\lambda_{c}$ but thin enough such
that spin-flip processes can be disregarded. We also assume that spin-flip
in the central normal metal node is negligible. We are in the diffuse
scattering regime, so $\hat{f}_{N}$ is isotropic and constant in space.

Referring back to Eq.~(\ref{currentoperator}), we need now quantum
statistical averages $\langle
a_{Am\alpha}^{\dagger}(E)a_{Bn\beta}(E^{\prime})\rangle=\delta_{AB}%
\delta_{mn}\delta(E-E^{\prime})f_{A}^{\beta\alpha}(E)$, where $a_{Bn\beta}$
is the annihilation operator for electrons moving in normal metal $A$ ($A=L,R
$ or $N$) towards one of the ferromagnets, and $f_{A}^{\beta\alpha}$ is the $%
\beta\alpha$-component of the $2\times2$ semiclassical distribution matrix $%
\hat{f}_{A}$ in spin space. For the reservoirs ($A=L$ or $R$), we simply
have $f_{A}^{\beta\alpha}=\delta _{\beta\alpha}f(E-\mu_{A})$. In contrast,
in the central node the spin accumulation is not necessarily parallel to the
spin quantization axis in either of the ferromagnets, meaning that
non-diagonal $\left( \beta\neq\alpha\right) $ terms in the distribution
matrix do not vanish. The average charge current flowing from the right into
ferromagnet $F_{1}$ can then be expressed by the generalized LB expressions%
\cite{epjb22,Brataas-review} 
\begin{align}
\langle I_{c,1R}\rangle & =\frac{e}{h}\int dE\left[g_{1}^{\uparrow}(f_{cN}+%
\mathbf{f}_{sN}\cdot\mathbf{m}_{1}-f_{L})\right.  \notag \\
& \hspace{1.8cm}\left.+g_{1}^{\downarrow}(f_{cN}-\mathbf{f}_{sN}\cdot\mathbf{%
m}_{1}-f_{L})\right],   \label{avgcharge}
\end{align}
whereas the average spin current reads 
\begin{align}
\langle\mathbf{I}_{s,1R}\rangle & =\frac{1}{4\pi}\int dE\left\{\mathbf{m}_{1}%
\left[g_{1}^{\uparrow}(f_{cN}+\mathbf{f}_{sN}\cdot\mathbf{m}%
_{1}-f_{L})\right.\right.  \notag \\
& \hspace{-1cm} \left.-g_{1}^{\downarrow}(f_{cN}-\mathbf{f}_{sN}\cdot\mathbf{%
m}_{1}-f_{L})\right]+2\mathrm{Re}g_{1R}^{\uparrow\downarrow}\mathbf{m}%
_{1}\times(\mathbf{f}_{sN}\times\mathbf{m}_{1})  \notag \\
&\hspace{-1cm}\left.+2\mathrm{Im}g_{1R}^{\uparrow\downarrow}\mathbf{f}%
_{sN}\times\mathbf{m}_{1}\right\}.   \label{avgspin}
\end{align}
Here $g_{1}^{\alpha}$ is the spin-dependent dimensionless conductance of $%
F_{1}$ and $g_{1R}^{\uparrow\downarrow}$ is the mixing conductance of the
interface between $F_{1}$ and the middle normal metal. The average charge
current and the component of the spin current polarized along the
magnetization are conserved through the ferromagnet. Hence $\langle
I_{c,1L}\rangle=-\langle I_{c,1R}\rangle$ and $\langle\mathbf{I}%
_{s,1L}\rangle\cdot\mathbf{m}_{1}=-\langle\mathbf{I}_{s,1R}\rangle\cdot%
\mathbf{m}_{1}$. The transverse spin current is absorbed in the ferromagnet,
leading to 
\begin{align}
\langle\mathbf{I}_{s,1L}\rangle & =\frac{1}{4\pi}\int dE\,\mathbf{m}_{1}%
\left[g_{1}^{\uparrow}(f_{L}-f_{cN}-\mathbf{f}_{sN}\cdot\mathbf{m}%
_{1})\right.  \notag \\
& \left.\hspace{2cm}-g_{1}^{\downarrow}(f_{L}-f_{cN}+\mathbf{f}_{sN}\cdot%
\mathbf{m}_{1})\right].   \label{avgspin2}
\end{align}
Similar expressions hold for the currents evaluated on the left and right
sides of $F_{2}$. In order to keep the expressions simple we adopt from now
on the parameters $g_{1}^{\alpha}=g_{2}^{\alpha}=g^{\alpha}$, and $%
g_{1L}^{\uparrow\downarrow}=g_{1R}^{\uparrow\downarrow}=g_{2L}^{\uparrow%
\downarrow}=g_{2R}^{\uparrow\downarrow}=g^{\uparrow\downarrow}$.

Since spin-flip processes are disregarded, both charge and spin are
conserved on the middle normal metal node: 
\begin{align}
\langle I_{c,1R}\rangle+\langle I_{c,2L}\rangle & =0 \\
\langle\mathbf{I}_{s,1R}\rangle+\langle\mathbf{I}_{s,2L}\rangle & =0 
\label{spinconservation}
\end{align}
Eqs.~(\ref{avgcharge})-(\ref{spinconservation}) come down to four equations
for the four unknown components of the distribution matrix $\hat{f}_{N}$ as
a function of the angle $\theta=\cos^{-1}(\mathbf{m}_{1}\cdot\mathbf{m}_{2})$
and the applied voltage $U=(\mu_{L}-\mu_{R})/e.$ Eq.~(\ref{avgcharge}) then
yields $\langle I_{c,1L}\rangle=-\langle I_{c,1R}\rangle=\langle
I_{c,2L}\rangle=-\langle I_{c,2R}\rangle\equiv I_{c}=G_{v}U,$ where\cite%
{Brataas-review} 
\begin{equation}
G_{v}=\frac{e^{2}g}{2h}\left( 1-P^{2}\frac{1-\mathrm{cos}\theta }{1-\mathrm{%
cos}\theta+\eta+\eta\mathrm{cos}\theta}\right)   \label{IsfaT}
\end{equation}
is the spin valve conductance with material parameters $g=g^{\uparrow
}+g^{\downarrow}$, $P=(g^{\uparrow}-g^{\downarrow})/g$, and $\eta
=2g^{\uparrow\downarrow}/g$ .

\subsection{Current noise}

\label{CurrentNoiseSpinValve}

We combine spin and charge current fluctuations, e.g., $\Delta{I}_{c,1R}(t)$
and $\Delta\mathbf{I}_{s,1R}(t)$, respectively, on the right side of $F_{1},$
into a $2\times2$ matrix in spin space: 
\begin{equation}
\Delta\hat{I}_{1R}(t)=\hat{1}\Delta{I}_{c,1R}(t)-(2e/\hbar)\hat{\boldsymbol{%
\sigma }}\cdot\Delta\mathbf{I}_{s,1R}(t)\,.
\end{equation}
Since we focus on the zero frequency noise, instantaneous charge and spin
conservation in the central node may be assumed, i.e. 
\begin{equation}
\Delta\hat{I}_{1R}(t)+\Delta\hat{I}_{2L}(t)=0,   \label{fluctcons}
\end{equation}
which requires that the distribution matrix in the node fluctuates. The
current fluctuations can then be written 
\begin{equation}
\Delta\hat{I}_{1R(2L)}(t)=\delta\hat{I}_{1R(2L)}(t)+\frac{\partial\langle 
\hat{I}_{1R(2L)}\rangle}{\partial\hat{f}_{N}}\delta\hat{f}_{N}(t), 
\label{totalfluct}
\end{equation}
where $\delta\hat{f}_{N}(t)$ are the fluctuations of the distribution
matrix, and $\delta\hat{I}_{1R(2L)}(t)$ are the intrinsic fluctuations [when 
$\delta\hat{f}_{N}(t)=0$], coinciding with the fluctuations calculated for
single ferromagnets in the previous section. Expression (\ref{totalfluct})
applies also to the current fluctuations evaluated on the left side of
ferromagnet $F_{1}$ and the right side of ferromagnet $F_{2}$. In the
following, we focus on thermal current noise, recalling from Sec. \ref%
{recipe} that for typical voltage drops in spin valves, shot noise is only
important at low temperatures.

From Eqs. (\ref{avgcharge}), (\ref{avgspin}), (\ref{fluctcons}) and (\ref%
{totalfluct}) and results from Sec. \ref{SingleFerromagnet}, we can evaluate
the charge and spin current fluctuations in the spin valve. The correlator $%
S_c(0)=\int d(t-t^{\prime})\langle\Delta{I}_{c}(t)\Delta{I}%
_{c}(t^{\prime})\rangle$ of the charge current fluctuations is simply
related to the conductance (\ref{IsfaT}) by the following
configuration-dependent FDT: 
\begin{equation}
S_c(\omega=0;\theta)=2k_{B}TG_{v}(\theta).
\end{equation}
In the low-frequency regime considered here, charge current noise is the
same anywhere in the spin valve. $G_{v}$ can vary easily by a factor of two
as a function of $\theta$, which corresponds to the same variation in noise
power. Resistance noise via magnetization fluctuations is an additional
source of electric noise that is treated below

The spin current correlator $\langle\Delta I_{s_{i},A}(t)\Delta
I_{s_{j},B}(t^{\prime})\rangle$, where $i$ and $j$ denote Cartesian
components and $A(B)=1L,1R,2L$ or $2R$, can be found analogously. Since spin
current is not conserved at the ferromagnetic interfaces, the spin current
correlator depends on the location in the spin valve and is not directly
observable. We therefore proceed to evaluate the magnetization fluctuations
caused by the spin current noise in the next subsection.

\subsection{Magnetization noise and damping}

The current-induced stochastic field acting on $F_{1}$ follows from the spin
current fluctuations as explained in Sec. \ref{recipe}. Here we discuss this
field and, by using the FDT, the corresponding Gilbert damping enhancement
in spin valves. In order to keep the algebra manageable, we focus on the
most relevant parallel, antiparallel and perpendicular configurations $%
\left( \mathrm{cos}\theta=0,\pm1\right) $. The mixing conductances are taken
to be identical for all four F$|$N-interfaces. In the semiclassical
approach, intrinsic current fluctuations are not correlated across the node,
implying that $\langle\delta I_{s_{i},1R}(t)\delta
I_{s_{j},2L}(t^{\prime})\rangle=0$.

\subsubsection{Parallel configuration}

For the parallel (P) magnetic configuration, $\mathbf{m}_{1}\cdot\mathbf{m}%
_{2}=1$, the thermal spin current-induced stochastic magnetic field in
ferromagnet $F_{1}$ reads 
\begin{equation}
\langle h_{i}^{(\mathrm{th})}(t)h_{j}^{(\mathrm{th})}(t^{\prime})\rangle
_{P}=2k_{B}T\frac{\alpha_{sv}}{\gamma_{0}M_{s}V}\delta_{ij}\delta(t-t^{%
\prime })
\end{equation}
where $i,j$ label vector components perpendicular to the magnetization, and 
\begin{equation}
\alpha_{sv}=\frac{3\gamma_{0}\hbar\mathrm{Re}g^{\uparrow\downarrow}}{8\pi
M_{s}V}.   \label{PincreaseAlpha}
\end{equation}
By the FDT, $\alpha_{sv}$ is identical to the spin-pumping enhancement of
the Gilbert damping of the $F_{1}$ magnetization. This can be checked by
following the steps outlined for a single ferromagnet, Eqs.~(\ref{FDT2})-(%
\ref{suscmatrix}). A possible exchange coupling between the ferromagnets
modifies the dynamics via $\mathbf{H}_{\mathrm{eff}}$ in the LLG equation,
but does not affect the stochastic field and Gilbert damping.

The field correlator and damping for the parallel configuration is reduced
by a factor $3/4$ compared with (\ref{advertisedResult}) for the single
ferromagnet sandwiched by normal metals. This result may be found also in a
more direct way: Using Eqs.~(\ref{spinpumping1}) and (\ref{avgspin}) we can
compute the net spin angular momentum leaving each of the ferromagnets when
the magnetizations are slightly out of equilibrium, and by conservation of
angular momentum infer the corresponding enhancement of the Gilbert damping
constant. The factor $3/4$ follows from the diffuse/chaotic nature of the
node: Half of the spin current that is pumped into the node is reflected
back and reabsorbed by $F_{1}.$

One subtle point needs to be noted in this discussion: When the F-N
interfaces are nearly transparent, the interfacial conductance parameters
from scattering theory should be corrected for spurious so-called Sharvin
conductances (See Sec. II.B. of Ref. \onlinecite{Tserkovnyakreview}). In
practice, this will correct (\ref{PincreaseAlpha}) only by a numerical
prefactor close to one.

\subsubsection{Antiparallel configuration}

For the antiparallel (AP) configuration ($\mathbf{m}_{1}\cdot\mathbf{m}%
_{2}=-1$), 
\begin{equation}
\langle h_{i}^{(\mathrm{th})}(t)h_{j}^{(\mathrm{th})}(t^{\prime})\rangle
_{AP}=\langle h_{i}^{(\mathrm{th})}(t)h_{j}^{(\mathrm{th})}(t^{\prime})%
\rangle_{P},
\end{equation}
\textit{i.e}., the current-induced noise and damping is the same as in the P
configuration. This result holds only when the imaginary part of the mixing
conductance is negligibly small.

\subsubsection{Perpendicular configuration}

When the $F_{2}$ magnetization is pinned along the $x$-direction and $%
\mathbf{m}_{1}$ points along the $z$-axis 
\begin{align}
\langle h_{x}^{(\mathrm{th})}(t)h_{x}^{(\mathrm{th})}(t^{\prime})\rangle_{%
\perp} & =2k_{B}T\frac{\alpha_{xx}^{\prime}}{\gamma _{0}M_{s}V}%
\delta(t-t^{\prime}),  \label{hPerp1} \\
\langle h_{y}^{(\mathrm{th})}(t)h_{y}^{(\mathrm{th})}(t^{\prime})\rangle_{%
\perp} & =2k_{B}T\frac{\alpha_{yy}^{\prime}}{\gamma _{0}M_{s}V}%
\delta(t-t^{\prime}),   \label{hPerp2}
\end{align}
where the subscript $\perp$ emphasizes that this is valid for the
perpendicular configuration, and, according to the FDT, 
\begin{align}
\alpha_{xx}^{\prime} & =\frac{3\gamma_{0}\hbar\mathrm{{Re}}%
g^{\uparrow\downarrow}}{8\pi M_{s}V}\,,  \notag \\
\alpha_{yy}^{\prime} & =\frac{\gamma_{0}\hbar\mathrm{{Re}}%
g^{\uparrow\downarrow}}{4\pi M_{s}V}\left[ 2-\frac{\eta(2-P^{2}+2\eta )}{%
2(1+\eta)(1-P^{2}+\eta)}\right]  \label{Gilbert:perpendicular}
\end{align}
is the spin pumping-induced enhancement of the Gilbert damping. The cross
correlators $\langle h_{x}^{(\mathrm{th})}(t)h_{y}^{(\mathrm{th}%
)}(t^{\prime})\rangle_{\perp}=\langle h_{y}^{(\mathrm{th})}(t)h_{x}^{(%
\mathrm{th})}(t^{\prime})\rangle_{\perp}=0$. In non-collinear spin valves,
the noise correlators and the Gilbert damping are therefore tensors. This
can be accommodated by the LLG equation for $\mathbf{m}_{1}$ by a damping
torque $\mathbf{m}_{1}\times\overleftrightarrow {\alpha}d\mathbf{m}_{1}/dt$,
where the Gilbert damping tensor (in the plane perpendicular to the
magnetization) reads: 
\begin{equation}
\overleftrightarrow{\alpha}=\left( 
\begin{array}{cc}
\alpha_{0}+\alpha_{xx}^{\prime} & 0 \\ 
0 & \alpha_{0}+\alpha_{yy}^{\prime}%
\end{array}
\right) .
\end{equation}
Note that the damping tensor must be written inside the cross product in the
damping torque to ensure that the LLG equation preserves the length of the
unit magnetization vector.

In our evaluation of the Gilbert damping (\ref{Gilbert:perpendicular}), we
have assumed that the outer left and right reservoirs have a fixed chemical
potential which allows charge current fluctuations into the reservoirs. This
is valid when the reservoirs are connected to external circuit elements with
sufficiently long $RC$-times compared to the FMR precession period. In the
opposite limit, when the reservoirs are fully decoupled from other circuit
elements, charge current into the reservoirs must vanish at any time, and
the chemical potentials fluctuate. This regime was considered in Ref. %
\onlinecite{DynamicStiffness2003} with the result 
\begin{align}
\alpha_{xx}^{\prime} & =\frac{3\gamma_{0}\hbar\mathrm{{Re}}%
g^{\uparrow\downarrow}}{8\pi M_{s}V}\,,  \notag \\
\alpha_{yy}^{\prime} & =\frac{\gamma_{0}\hbar\mathrm{{Re}}%
g^{\uparrow\downarrow}}{4\pi M_{s}V}\left[ 2-\frac{\eta}{1-P^{2}+\eta}\right]
\end{align}

\subsection{Resistance noise}

The fluctuations of the magnetization vector can be calculated by the LLG
equation that incorporates the stochastic fields. Fluctuations in the
magnetic configuration affect the electrical resistance that depends on the
dot product $\mathbf{m}_{1}\cdot\mathbf{m}_{2}$. Resistance noise is an
important issue for application of spin-valve read heads.\cite{Smith1}
Covington \textit{et al.}\cite{Covington} measured resistance noise in
current-perpendicular-to-the-plane (CPP) spin valves, which are considered
an alternative for the conventional current-in-the-plane spin valve read
heads. We focus here on the zero-frequency resistance noise 
\begin{equation}
S_{R}(\omega=0)=\int d(t-t^{\prime})\langle\Delta R(t)\Delta R(t^{\prime
})\rangle,
\end{equation}
where $\Delta R(t)$ is the time dependent deviation of the resistance from
the time-averaged value.

Resistance noise can be measured e.g. as voltage noise for constant current
bias or as current noise for a constant voltage bias. The resistance noise
comes on top of the Johnson-Nyquist noise discussed in Sec \ref%
{CurrentNoiseSpinValve} and in Ref. \onlinecite{Xiao}. We find that at relatively
high current densities, the magnetization-induced noise can be the dominant
contribution to the electric noise. The current densities considered are not
so high that shot noise dominates over Johnson-Nyquist noise, consistent
with our assumption that shot noise may be neglected.

In the following, we derive the resistance noise in the parallel,
antiparallel and perpendicular configurations. Recall that the magnetization
in ferromagnet $F_{2}$ is assumed pinned. The analysis of resistance noise
in the case of two fluctuating magnetizations is left for the next section.

\subsubsection{Parallel configuration}

The total stochastic field in $F_{1}$ causes fluctuations $\delta \mathbf{m}%
_{1}(t)=\mathbf{m}_{1}(t)-\langle\mathbf{m}_{1}\rangle$ relative to its
time-averaged equilibrium value $\langle\mathbf{m}_{1}\rangle$. For the
parallel configuration $\langle\mathbf{m}_{1}\rangle=\mathbf{m}_{2}$, such
that the dot product of the magnetizations is $\cos\theta=\mathbf{m}_{1}\cdot%
\mathbf{m}_{2}=1-\delta\mathbf{m}_{1}^{2}/2$, with $\theta$ the angle
between the magnetization directions. For small fluctuations we can expand
the resistance to first order in $\delta\mathbf{m}_{1}^{2}$ 
\begin{equation}
R(\mathbf{m}_{1}\cdot\mathbf{m}_{2})\approx R(1)-\frac{1}{2}\delta \mathbf{m}%
_{1}^{2}\frac{\partial R(1)}{\partial\cos\theta},
\end{equation}
such that the resistance noise correlator becomes 
\begin{align}
\langle\Delta R(t)\Delta R(t^{\prime})\rangle_{P} & =\langle R(t)R(t^{\prime
})\rangle_{P}-\langle R(t)\rangle_{P}\langle R(t^{\prime})\rangle _{P} 
\notag \\
& \hspace{-2cm}=\frac{1}{4}\left( \frac{\partial R(1)}{\partial\cos\theta}%
\right) ^{2}\left[\langle\delta\mathbf{m}_{1}^{2}(t)\delta\mathbf{m}%
_{1}^{2}(t^{\prime })\rangle_{P}\right.  \notag \\
&\hspace{1cm}\left.-\langle\delta\mathbf{m}_{1}^{2}(t)\rangle_{P}\langle%
\delta\mathbf{m}_{1}^{2}(t^{\prime})\rangle_{P}\right],
\end{align}
where the brackets denote statistical averaging around the parallel
configuration. Assuming that the stochastic fields are Gaussian distributed,
so are the fluctuations of the magnetization vectors, since the
magnetization is a linear function of the stochastic fields. We may then
employ Wick's theorem,\cite{Wick} according to which fourth order moments of
the fluctuations can be expressed in terms of the sum of products of second
order moments. We then arrive at 
\begin{align}
\langle\Delta R(t)\Delta R(t^{\prime})\rangle_{P} =& \frac{1}{2}\left( \frac{%
\partial R(1)}{\partial\cos\theta}\right) ^{2}  \notag \\
& \times\sum_{ij}\langle\delta m_{1,i}(t)\delta
m_{1,j}(t^{\prime})\rangle_{P}^{2},   \label{ResistanceNoiseP}
\end{align}
where $i$ and $j$ denote Cartesian components. From Eq.~(\ref{IsfaT}) we
find 
\begin{equation}
\frac{\partial R(1)}{\partial\cos\theta}=-\frac{hP^{2}}{e^{2}g\eta }. 
\label{diffP}
\end{equation}
Since the magnetization fluctuations are small, we may disregard their
longitudinal component, whereas the correlator of the transverse
fluctuations can be computed by the LLG equation.

We use the coordinate system in Fig. \ref{spinvalve} with interfaces in the $%
xz$-plane. The LLG equation reads 
\begin{align}
\frac{d\mathbf{m}_{1}}{dt}=&-\gamma_{0}\mathbf{m}_{1}\times\lbrack \mathbf{H}%
_{\mathrm{eff}}+\mathbf{h}(t)]  \notag \\
&+(\alpha_{0}+\alpha_{sv})\mathbf{m}_{1}\times\frac{d\mathbf{m}_{1}}{dt}, 
\label{LLGsv}
\end{align}
where the total stochastic field $\mathbf{h}(t)=\mathbf{h}^{(0)}(t)+\mathbf{h%
}^{\mathrm{(th)}}(t)$ includes both the intrinsic field $\mathbf{h}^{(0)}(t)$
(see section \ref{SingleFerromagnet}) and the current induced field $\mathbf{%
h}^{\mathrm{(th)}}(t)$ from the previous section. $\alpha_{0}$ and $%
\alpha_{sv}$ are the corresponding Gilbert damping parameters. The effective
field $\mathbf{H}_{\mathrm{eff}}=\mathbf{H}_{0}+\mathbf{H}_{a}+\mathbf{H}%
_{d}+\mathbf{H}_{e}$ contains the external field $\mathbf{H}_{0}$, the
in-plane anisotropy field $\mathbf{H}_{a}$, the out-of-plane demagnetizing
field $\mathbf{H}_{d}$, and the sum of dipolar and exchange fields $\mathbf{H%
}_{e}$. The external and anisotropy fields are both taken along the $z$%
-axis. We parametrize these fields by $\omega_{0}$ and $\omega_{a}$ as $%
\gamma\mathbf{H}_{0}=\omega_{0}\mathbf{z}$ and $\gamma\mathbf{H}%
_{a}=\omega_{a}(\mathbf{m}_{1}\cdot\mathbf{z})\mathbf{z}$. The demagnetizing
field is directed normal to the plane, \textit{i.e}. along the $y$-axis,
such that $\gamma\mathbf{H}_{d}=-\omega _{d}(\mathbf{m}_{1}\cdot\mathbf{y})%
\mathbf{y}$ thereby introducing the parameter $\omega_{d}$. The dipolar and
exchange couplings are described in terms of a Heisenberg coupling $-J%
\mathbf{m}_{1}\cdot\mathbf{m}_{2}$, which favors a parallel magnetic
configuration for $J>0$ and an antiparallel one for $J<0$. This translates
into the field $\gamma\mathbf{H}_{e}=\omega_{e}\mathbf{m}_{2}$, where $%
\omega_{e}=\gamma J/M_{s}d$.

In the P configuration $\langle\mathbf{m}_{1}\rangle$ is aligned with the
pinned $\mathbf{m}_{2}$ in the $+z$ direction, which can always be enforced
by a sufficiently strong external field. Linearizing the LLG equation in the
amplitude of the transverse fluctuations $\delta\mathbf{m}(t)\approx\delta {m%
}_{x}(t)\mathbf{x}+\delta{m}_{y}(t)\mathbf{y}$, we find the magnetization
noise correlator 
\begin{equation}
\langle\delta m_{i}(t)\delta m_{j}(t^{\prime})\rangle_{P}=\frac{\gamma
_{0}k_{B}T\alpha}{\pi M_{s}V}\int d\omega e^{-i\omega(t-t^{\prime})}U_{ij}, 
\label{dmdmP}
\end{equation}
by using the correlators of the stochastic fields. Here 
\begin{align}
U_{xx} & =\frac{[\omega^{2}+(\omega_{t}+\omega_{d})^{2}]}{%
[\omega^{2}-\omega_{t}(\omega_{t}+\omega_{d})]^{2}+\omega^{2}\alpha^{2}(2%
\omega _{t}+\omega_{d})^{2}}, \\
U_{xy} & =\frac{-i\omega(2\omega_{t}+\omega_{d})}{[\omega^{2}-\omega
_{t}(\omega_{t}+\omega_{d})]^{2}+\omega^{2}\alpha^{2}(2\omega_{t}+\omega
_{d})^{2}}, \\
U_{yy} & =\frac{(\omega^{2}+\omega_{t}^{2})}{[\omega^{2}-\omega_{t}(%
\omega_{t}+\omega_{d})]^{2}+\omega^{2}\alpha^{2}(2\omega_{t}+\omega_{d})^{2}}%
, \\
U_{yx} & =-U_{xy},
\end{align}
with $\alpha=\alpha_{0}+\alpha_{sv}$ and $\omega_{t}=\omega_{0}+\omega_{a}+%
\omega_{e}$. The above expressions hold for small damping, \textit{i.e}., $%
\alpha_{0}^{2},\alpha_{sv}^{2}\ll1$. The zero-frequency resistance noise $%
S_{P}(0)=\int d(t-t^{\prime})\langle\Delta R(t)\Delta
R(t^{\prime})\rangle_{P}$ is obtained by inserting Eq.~(\ref{dmdmP}) into
Eq. (\ref{ResistanceNoiseP}): 
\begin{align}
S_{P}(0) =& \frac{1}{\pi}\left( \frac{hP^{2}}{e^{2}g\eta}\right) ^{2}\left( 
\frac{\gamma_{0}k_{B}T\alpha}{M_{s}V}\right) ^{2}  \notag \\
& \times\int d\omega(U_{xx}^{2}+U_{yy}^{2}-2U_{xy}^{2}).   \label{Spfinal}
\end{align}
To gain insight into this rather complicated expression, it is convenient to
make some simplifications. Although the demagnetizing field, which serves to
stabilize the magnetization in the plane of the film, is important to get
the right magnitude of the noise, we can gain physical understanding by
disregarding it. Setting $\omega_{d}=0$, we find 
\begin{equation}
S_{P}(0)=\left( \frac{\gamma_{0}k_{B}T}{M_{s}V}\right) ^{2}\left( \frac{%
hP^{2}}{e^{2}g\eta}\right) ^{2}\frac{1}{\omega_{t}^{3}\alpha }. 
\label{SpfinalSimpl}
\end{equation}
Obviously, the resistance noise strongly depends on the parameter $\omega_{t}
$. The external and anisotropy fields stabilize the magnetization, hence
lowering the noise. The dipolar and exchange field either stabilizes or
destabilizes the magnetization, depending on the sign of the coupling
constant $J$. We observe that the Gilbert damping also strongly affects the
resistance noise. The resistance noise decreases with increasing damping,
because the suppression of the magnetic susceptibility by a large alpha
turns out to be more important than the FDT-motivated increase of the
stochastic field noise. Since $\alpha_{sv}$ can be of the same order as $%
\alpha_{0}$\cite{Tserkovnyakreview}, the importance of spin current noise
and spin pumping is evident.

When a constant voltage bias is applied, the resistance noise causes current
noise. At sufficiently small bias, the Johnson-Nyquist current noise (Sec.~%
\ref{CurrentNoiseSpinValve}) always wins. However, at relatively high
current densities, the effects of the resistance noise are very significant.
That noise may be important for the next generation magnetoresistive spin
valve read heads.\cite{Smith1} For a quantitative comparison, which depends
on many material parameters, it is important to use Eq.~(\ref{Spfinal}) and
not Eq.~(\ref{SpfinalSimpl}), since the demagnetizing field has a large
effect on the magnitude of the magnetization-induced noise. The
magnetization-induced noise is most prominent for small structures, since
the ratio of Johnson-Nyquist noise to magnetization-induced noise scales
with the volume of the ferromagnet.

\subsubsection{Antiparallel configuration}

When $J<0$, the dipolar and exchange coupling favors an AP configuration ($%
\langle\mathbf{m}_{1}\rangle=-\mathbf{m}_{2}$) at zero external magnetic
field. Following the recipe of the previous subsection, we find a resistance
noise%
\begin{align}
\langle\Delta R(t)\Delta R(t^{\prime})\rangle_{AP} =& \frac{1}{2}\left( 
\frac{\partial R(-1)}{\partial\cos\theta}\right) ^{2}  \notag
\label{ResistanceNoiseAP} \\
& \times\sum_{ij}\langle\delta m_{1,i}(t)\delta
m_{1,j}(t^{\prime})\rangle_{AP}^{2},
\end{align}
where the sensitivity of the resistance to the fluctuations is 
\begin{equation}
\frac{\partial R(-1)}{\partial\cos\theta}=-\frac{hP^{2}\eta}{%
e^{2}g(1-P^{2})^{2}}.
\end{equation}
Using the magnetization noise correlators from the linearized Eq. (\ref%
{LLGsv}), the zero frequency resistance noise becomes 
\begin{align}
S_{AP}(0) =& \int d(t-t^{\prime})\langle\Delta R(t)\Delta R(t^{\prime
})\rangle_{AP}  \notag \\
=& \frac{1}{\pi}\left( \frac{hP^{2}\eta}{e^{2}g(1-P^{2})^{2}}\right)
^{2}\left( \frac{\gamma_{0}k_{B}T\alpha}{M_{s}V}\right) ^{2}  \notag \\
& \times\int d\omega(V_{xx}^{2}+V_{yy}^{2}-2V_{xy}^{2}),   \label{Sapfinal}
\end{align}
where $V_{ij}=U_{ij}(\omega_{t}\rightarrow\omega_{s})$ with $%
\omega_{s}=\omega_{a}-\omega_{e}$ (recall that $\omega_{e}<0$). Again
disregarding the demagnetizing field strongly simplifies the expression: 
\begin{equation}
S_{AP}(0)=\left( \frac{\gamma_{0}k_{B}T}{M_{s}V}\right) ^{2}\left( \frac{%
hP^{2}\eta}{e^{2}g(1-P^{2})^{2}}\right) ^{2}\frac{1}{\omega_{s}^{3}\alpha}.
\end{equation}
As expected, the resistance noise decreases with increasing $\omega_{s}$.
The anisotropy, dipolar and exchange fields stabilizes the magnetization,
playing a role similar to that of the external field in the P configuration.
The Gilbert damping enters in the same way as for the P configuration.

Except for the prefactor that reflects the sensitivity of the resistance to
the magnetization fluctuations, $S_{P}(0)$ and $S_{AP}(0)$ are very similar.
For the special case $\omega_{t}=\omega_{s}$, 
\begin{equation}
\frac{S_{P}}{S_{AP}}=\frac{(1-P^{2})^{4}}{\eta^{4}}.
\end{equation}
For, e.g., $P=0.7$ and $\eta=1$, this becomes $S_{P}/S_{AP}\approx6\%$
showing that the difference in noise level between the P and AP
configurations can be substantial.

This asymmetry in the noise level between the P and AP configurations is
consistent with the the experimental results of Covington et al. on nearly
cylindrical multilayer pillars.\cite{Covington2} In these experiments the
magnetizations were aligned parallel when the external magnetic field
reached about 1500 Oe. Although we treat spin valves with two ferromagnetic
films and Covington et al. dealt with multilayers of 4-15 magnetic films, it
is likely that the difference between the noise properties of bilayers and
multilayers is small, as the only local structural difference is the number
of neighboring ferromagnets. This assertion is supported by the experiments
by Covington et al. that did not reveal strong differences for nanopillars
ranging from 4-15 layers.

\subsubsection{Perpendicular configuration}

We now investigate the perpendicular state $\langle\mathbf{m}_{1}\rangle
\cdot\mathbf{m}_{2}=0$, assuming that $\mathbf{m}_{2}$ now has been pinned
in the $x$-direction, whereas $\mathbf{m}_{1}$ is on average parallel to the 
$z$ axis, as before. In the following we assume that the interlayer exchange
and dipolar coupling are negligibly small, since otherwise the algebra and
expressions become awkward.

Expanding the resistance to first order in the fluctuations $\delta \mathbf{m%
}_{1}$, we find in this case 
\begin{equation}
\langle\Delta R(t)\Delta R(t^{\prime})\rangle_{\perp}=\left( \frac{\partial
R(0)}{\partial\cos\theta}\right)^{2}\langle\delta m_{1x}(t)\delta
m_{1x}(t^{\prime})\rangle.  \label{ResistancePerp}
\end{equation}
The magnetization fluctuations affect the resistance noise in the
perpendicular configuration to second order, unlike for the P and AP
configurations, in which the leading term was of fourth order. The
sensitivity of the resistance for this configuration is according to Eq. (%
\ref{IsfaT}) 
\begin{equation}
\frac{\partial R(0)}{\partial\cos\theta}=-\frac{4hP^{2}\eta}{%
e^{2}g(1+\eta-P^{2})^{2}},   \label{diffPerp}
\end{equation}
Linearizing Eq. (\ref{LLGsv}) and using the correlators Eqs. (\ref{hPerp1})
and (\ref{hPerp2}) for the stochastic field we find 
\begin{widetext}
\begin{equation}
	\langle\delta m_{1x}(t)\delta m_{1x}(t')\rangle = \frac{\gamma_0 k_BT}{\pi M_sV}\int d\omega e^{-i\omega(t-t')}
							\frac{\omega^2(\alpha_0+\alpha^{\prime}_{yy})+(\omega_p+\omega_d)^2(\alpha_0+\alpha^{\prime}_{xx})}
	{[\omega^2-\omega_p(\omega_p+\omega_d)]^2+\omega^2[\omega_p(2\alpha_0+\alpha_{xx}^{\prime}+\alpha_{yy}^{\prime})+\omega_d(\alpha_0+\alpha_{xx}^{\prime})]^2},
\end{equation}
\end{widetext}where $\omega_{p}=\omega_{0}+\omega_{c}$. We then arrive at
the zero-frequency resistance noise 
\begin{equation}
S_{\perp}(0)=\frac{2\gamma_{0}k_{B}T}{M_{s}V}\left( \frac{4hP^{2}\eta}{%
e^{2}g(1+\eta-P^{2})^{2}}\right) ^{2}\frac{\alpha_{0}+\alpha _{xx}^{\prime}}{%
\omega_{p}^{2}},
\end{equation}
quite different from that in the collinear configurations. In particular,
the damping appears here in the numerator and there is no dependence on the
demagnetizing field. Notice that since $S_{\perp}$ is quadratic in magnetic
fluctuations [see Eq.(\ref{ResistancePerp})], it becomes linear in
temperature, unlike $S_P$ and $S_{AP}$.

\subsection{Two identical ferromagnets\label{TIFM}}

We now investigate spin valves in which the ferromagnets are identical and
hence equally susceptible to fluctuations,\cite{forosPRB} focusing now only
on the P and AP configurations. The fluctuations of $F_{1}$ are $\delta 
\mathbf{m}_{1}(t)=\mathbf{m}_{1}(t)-\langle\mathbf{m}_{1}\rangle$ and those
of $F_{2}$ are $\delta\mathbf{m}_{2}(t)=\mathbf{m}_{2}(t)-\langle\mathbf{m}%
_{2}\rangle$. As before, we choose the $z$-axis so that the time-averaged
equilibrium values are $\langle\mathbf{m}_{1}\rangle=\langle\mathbf{m}%
_{2}\rangle=\mathbf{z}$ for the parallel configuration, and $\langle\mathbf{m%
}_{1}\rangle=-\langle\mathbf{m}_{2}\rangle=\mathbf{z}$ for the antiparallel.
The dot product of the magnetizations is $\mathbf{m}_{1}\cdot\mathbf{m}%
_{2}=\pm1\mp(\delta\mathbf{m}^{\mp})^{2}/2$, where the upper (lower) sign
holds for the P (AP) orientation and $\delta \mathbf{m}^{\mp}=\delta\mathbf{m%
}_{1}\mp\delta\mathbf{m}_{2}$. For small fluctuations, we can expand the
resistance to first order in $(\delta \mathbf{m}^{\mp})^{2}$, finding 
\begin{equation}
R(\mathbf{m}_{1}\cdot\mathbf{m}_{2})\approx R(\pm)\mp\frac{1}{2}(\delta%
\mathbf{m}^{\mp})^{2}\frac{\partial R(\pm1)}{\partial\cos\theta}.
\end{equation}
The resistance noise is then 
\begin{widetext}
\begin{align}
	\langle\Delta R(t)\Delta R(t')\rangle_{P/AP} & =  \langle R(t)R(t')\rangle_{P/AP}-\langle R(t) \rangle_{P/AP}\langle R(t')\rangle_{P/AP}	\nonumber\\
					      & =  \frac{1}{4}\left(\frac{\partial R(\pm 1)}{\partial\rm{cos}\theta}\right)^2
				\left[\langle(\delta\mathbf{m}^{\mp})^2(\delta\mathbf{m}^{\mp})^2\rangle_{P/AP}
			-\langle(\delta\mathbf{m}^{\mp})^2\rangle_{P/AP}\langle(\delta\mathbf{m}^{\mp})^2\rangle_{P/AP}\right],  	
\end{align}
\end{widetext}which by employing Wick's theorem becomes 
\begin{align}
\langle\Delta R(t)\Delta R(t^{\prime})\rangle_{P/AP} & =\frac{1}{2}\left( 
\frac{\partial R(\pm1)}{\partial\cos\theta}\right) ^{2}  \notag \\
& \times\sum_{ij}\langle\delta m_{i}^{\mp}(t)\delta m_{j}^{\mp}(t^{\prime
})\rangle_{P/AP}^{2}.  \label{ResistanceNoise}
\end{align}

Letting the subscripts $k$ and $l$ refer to ferromagnet $1$ or $2$, the LLG
equation in this case reads 
\begin{align}
\frac{d\mathbf{m}_{k}}{dt} & =-\gamma_{0}\mathbf{m}_{k}\times\lbrack \mathbf{%
H}_{\mathrm{eff}}+\mathbf{h}_{k}(t)]  \notag \\
& +(\alpha_{0}+\alpha_{sv})\mathbf{m}_{k}\times\frac{d\mathbf{m}_{k}}{dt}+%
\frac{\alpha_{sv}}{3}\mathbf{m}_{l}\times\frac{d\mathbf{m}_{l}}{dt}, 
\label{LLGsv2}
\end{align}
where the effective field $\mathbf{H}_{\mathrm{eff}}$ is now taken to be
equal for both ferromagnets. Due to current conservation, the ferromagnets
respective current-induced stochastic fields are not independent of each
other. With the spin current noise calculated in Sec.~\ref%
{CurrentNoiseSpinValve}, and following the recipe in Sec.~\ref{recipe}, we
find 
\begin{equation}
\langle h_{1,i}^{(\mathrm{th})}(t)h_{2,j}^{(\mathrm{th})}(t^{\prime})%
\rangle_{P}=-2k_{B}T\frac{\alpha_{sv}/3}{\gamma_{0}M_{s}V}%
\delta_{ij}\delta(t-t^{\prime})
\end{equation}
for the P configuration, and 
\begin{equation}
\langle h_{1,i}^{(\mathrm{th})}(t)h_{2,j}^{(\mathrm{th})}(t^{\prime})%
\rangle_{AP}=2k_{B}T\frac{\alpha_{sv}/3}{\gamma_{0}M_{s}V}%
\delta_{ij}\delta(t-t^{\prime}).
\end{equation}
for the AP configuration (as before $i,j$ label components perpendicular to
the magnetization direction). $\alpha_{sv}$ is defined in Eq. (\ref%
{PincreaseAlpha}). Naturally, the bulk fields $\mathbf{h}_{1}^{(0)}$ and $%
\mathbf{h}_{2}^{(0)}$ are uncorrelated. The last term in the LLG Eq. (\ref%
{LLGsv2}) represent the dynamic spin-exchange coupling:\cite%
{HeinrichBrataas,Tserkovnyakreview} It is the spin current pumped from
ferromagnet $l$ (see Sec.~\ref{SingleFerromagnet}) that is transmitted to
and subsequently absorbed by ferromagnet $k$. Since the normal metal node is
chaotic, this amounts to one third of the net total spin current pumped out
of ferromagnet $l$. This dynamic coupling was not present in spin valves in
which one magnetization is not moving at all.

By linearizing Eq. (\ref{LLGsv2}) in $\delta\mathbf{m}_{k}(t)$ we can
evaluate the desired magnetization noise correlators that are to be inserted
in Eq. (\ref{ResistanceNoise}). The zero-frequency resistance noise for the
P and AP configurations then respectively reads 
\begin{align}
S_{P}(0) =& \frac{1}{\pi}\left( \frac{hP^{2}}{e^{2}g\eta}\right) ^{2}\left( 
\frac{2\gamma_{0}k_{B}T}{M_{s}V}\right) ^{2}  \notag \\
& \times\int d\omega(Z_{xx}^{2}+Z_{yy}^{2}-2Z_{xy}^{2}).   \label{SpfinalII}
\end{align}
and 
\begin{align}
S_{AP}(0) =& \frac{1}{\pi}\left( \frac{hP^{2}\eta}{e^{2}g(1-P^{2})^{2}}%
\right) ^{2}\left( \frac{2\gamma_{0}k_{B}T}{M_{s}V}\right) ^{2}  \notag \\
& \times\int d\omega(X_{x}^{2}+X_{y}^{2}).   \label{SapfinalII}
\end{align}
Here 
\begin{align}
Z_{xx} & =\frac{\alpha_{t}[\omega^{2}+(\omega_{i}+\omega_{d})^{2}]}{%
[\omega^{2}-\omega_{i}(\omega_{i}+\omega_{d})]^{2}+\omega^{2}%
\alpha_{t}^{2}(2\omega_{i}+\omega_{d})^{2}}, \\
Z_{xy} & =\frac{-i\omega\alpha_{t}(2\omega_{i}+\omega_{d})}{[\omega
^{2}-\omega_{i}(\omega_{i}+\omega_{d})]^{2}+\omega^{2}\alpha_{t}^{2}(2%
\omega_{i}+\omega_{d})^{2}}, \\
Z_{yy} & =\frac{\alpha_{t}(\omega^{2}+\omega_{i}^{2})}{[\omega^{2}-\omega
_{i}(\omega_{i}+\omega_{d})]^{2}+\omega^{2}\alpha_{t}^{2}(2\omega_{i}+%
\omega_{d})^{2}},
\end{align}
and 
\begin{widetext}
\begin{eqnarray}
	X_x &=& \frac{\omega^2\alpha_s+(\omega_c+\omega_d)^2\alpha_t}
		 {[\omega^2+(\omega_c+\omega_d)(2\omega_e-\omega_c)]^2+\omega^2(2\omega_x\alpha_s-2\omega_c\alpha-\omega_d\alpha_t)^2}, \\
	X_y &=& \frac{\omega^2\alpha_s+\omega_c^2\alpha_t}
		 {[\omega^2+\omega_c(2\omega_e-\omega_c-\omega_d)]^2+\omega^2(2\omega_x\alpha_s-2\omega_c\alpha-\omega_d\alpha_s)^2}.
\end{eqnarray}
\end{widetext} For convenience, we defined $\alpha_{s}=\alpha_{0}+2\alpha
_{sv}/3$, $\alpha_{t}=\alpha_{0}+4\alpha_{sv}/3$, $\alpha=\alpha_{0}+%
\alpha_{sv}$ (note the difference between $\alpha$, $\alpha_{s}$ and $%
\alpha_{t}$), and $\omega_{i}=\omega_{0}+\omega_{a}+2\omega_{x}$. The above
expressions hold for small damping, \textit{i.e}., $\alpha_{0}^{2},\alpha
_{sv}^{2}\ll1$.

\begin{figure}[ptb]
\includegraphics[width=0.45\textwidth]{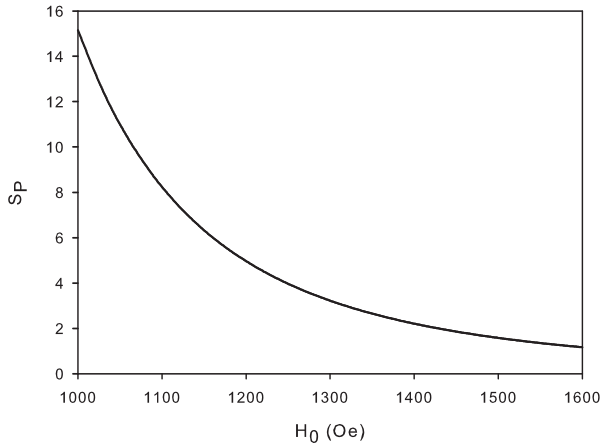}
\caption{The resistance noise in the P configuration as a function of the
externally applied magnetic field, given in units of $(10^{-7}/\protect\pi%
)(hP^2/e^2g\protect\eta)^2(2\protect\gamma_0 k_BT/M_sV)^2$. The parameters
used are $\protect\alpha_0=\protect\alpha_{sv}=0.01$, $\protect\omega_c/%
\protect\gamma_0=\protect\omega_d/\protect\gamma_0=100$~Oe and $J=-0.10$%
~erg/cm$^2$.}
\label{SpExtfield}
\end{figure}

Compared to the results in the previous section, we see that Eq. (\ref%
{SpfinalII}) is similar to Eq. (\ref{Spfinal}), whereas Eq. (\ref{SapfinalII}%
) differs considerably from Eq. (\ref{Sapfinal}). This is due to the static
dipolar and exchange couplings, and the dynamic spin-exchange coupling,
whose effects on the noise are modified by the presence of the second
fluctuating ferromagnet. In particular, the latter coupling causes the
Gilbert damping constant to enter Eqs. (\ref{SpfinalII}) and (\ref%
{SapfinalII}) differently. Eq. (\ref{SpfinalII}) decreases with the external
field and Eq. (\ref{SapfinalII}) decreases with the dipolar and exchange
coupling, as expected, and as shown in Figs. \ref{SpExtfield} and \ref%
{SapExchange}. The noise level is in general higher when both ferromagnets
fluctuate, than when only one does.

\begin{figure}[ptb]
\includegraphics[width=0.45\textwidth]{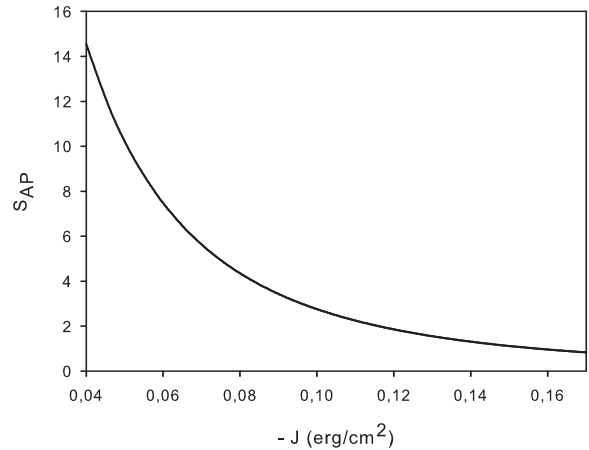}
\caption{The resistance noise in the AP configuration as a function of the
dipolar and exchange coupling between the ferromagnets, given in units of $%
(10^{-7}/\protect\pi)[hP^2\protect\eta/e^2g(1-P^2)^2]^2(2\protect\gamma_0
k_BT/M_sV)^2$. The parameters used are $\protect\alpha_0=\protect\alpha%
_{sv}=0.01$, $\protect\omega_c/\protect\gamma_0=\protect\omega_d/\protect%
\gamma_0=100$~Oe, and $\mathbf{H}_0=0$.}
\label{SapExchange}
\end{figure}

The resistance noise is governed by a number of material parameters.
Depending on these parameters, the noise level in the P configuration can
differ substantially from that in the AP configuration. Note that Eq. (\ref%
{SapfinalII}) reduces to that of Ref. \onlinecite{forosPRB} when the
demagnetizing field is disregarded, i.e., when $\omega_{d}\rightarrow0$,
whereas Eq. (\ref{SpfinalII}) does when $\omega_{a}\rightarrow0$ and $%
\omega_{d}\rightarrow-\omega_{a}$, since the external field in our earlier
work was perpendicular to the anisotropy field. The considerable difference
between $S_{P}(0)$ can $S_{AP}(0)$ in typical experimental spin-valve setups
can partly be explained by the dynamic exchange coupling .\cite{forosPRB}
However, also the sensitivity of the resistance to magnetic configuration
changes can be important, as shown in the previous section. The
demagnetizing field also significantly affects the numerical result for the
noise level since it stabilizes the magnetization, in both the P and AP
configurations.

\section{Conclusions}

Using scattering theory and magnetoelectronic circuit theory, we demonstrate
the effect of spin current fluctuations on the magnetization in
ferromagnetic multilayers. Via a fluctuating spin-transfer torque, the
current noise causes significantly enhanced magnetization noise, which in
spin valves is a function of the magnetic configuration. The noise is
related to the magnetization damping by the FDT, and can be experimentally
detected as resistance noise. The contribution from spin current noise to
resistance noise is considerable, and may be an issue for the next
generation magnetoresistive spin valve read heads.

\begin{acknowledgments}
We thank Mark Covington for sharing his results prior to publication and
Hans Joakim Skadsem for discussions. This work was supported by EC Contract
IST-033749 \textquotedblleft DynaMax\textquotedblright.
\end{acknowledgments}

%\bibliographystyle{unsrt}
%\bibliographystyle{plain}
%\bibliography{DiplomBibliografi}

\end{document}